\documentclass[twocolumn,english,prl,epsfig,rotate,showpacs,aps]{revtex4-2}
\usepackage[T1]{fontenc}
\usepackage[latin9]{inputenc}
\usepackage{color}
\usepackage{epsfig}
\usepackage{verbatim}
\usepackage{amsmath}
\usepackage{amsthm}
\usepackage{amssymb}
\usepackage{graphicx}
\usepackage{mathrsfs}

\usepackage{cancel}
\usepackage{esint}
\usepackage{microtype}
\usepackage{dcolumn}
\usepackage{bm}
\usepackage{amsfonts}
\usepackage[table]{xcolor} %

\makeatletter
\usepackage{babel}
\usepackage{braket}

\makeatother

\usepackage{babel}

\begin{document}

\renewcommand{\figurename}{Fig.}
\title{The Effect of the Non-Abelian Quantum Metric on Superfluidity} 
\author{Kai Chen$^{\chi}$}
\affiliation{Department of Physics and Texas Center for Superconductivity, University of Houston, Houston, TX 77204}
\author{Bishnu Karki}
\affiliation{Department of Physics and Texas Center for Superconductivity, University of Houston, Houston, TX 77204}
\author{Pavan Hosur$^{\dagger}$}
\affiliation{Department of Physics and Texas Center for Superconductivity, University of Houston, Houston, TX 77204}

\date{\today}

\begin{abstract}
The quantum geometric tensor, which encodes the full geometric information of quantum states in projective Hilbert space, plays a crucial role in condensed matter physics. In this work, we examine the effect of the non-Abelian quantum metric---the real part of the non-Abelian quantum geometric tensor---on the superfluid weight in time-reversal symmetric systems. For conventional $s$-wave pairing, we demonstrate that the superfluid weight includes a contribution proportional to the trace of the non-Abelian quantum metric. Notably, this contribution remains significant even when the total Chern number of a set of degenerate bands is zero and can exceed the conventional contribution, as confirmed using lattice models. \textit{Ab initio} density functional theory (DFT) calculations for MoS$_2$ and TiSe$_2$ further corroborate these findings, revealing that the non-Abelian quantum metric accounts for up to 20\% of the superfluid weight in MoS$_2$ and 50\% in TiSe$_2$. Our results provide new insights into the nontrivial relationship between the geometric properties of quantum states and superconductivity, opening avenues for further exploration in topological and superconducting materials.
\end{abstract}
\maketitle

\textit{Introduction}\textbf{---} The geometric properties of quantum states in projective Hilbert space have sparked conceptual revolutions in modern physics. One of the most important geometric features is the Berry connection, which underpins the theory of polarization \cite{Resta2007,spaldin2012beginner}. In an insulator, the Berry curvature integrated over the two-dimensional Brillouin zone (BZ), summed over all occupied bands, defines a topological invariant---the Chern number \cite{thouless1982quantized,hatsugai1993chern}. This invariant not only explains the quantization of Hall conductivity but also classifies insulators into distinct topological phases \cite{hasan2010colloquium,qi2011topological,ryu2010topological,moore2010birth,ando2013topological}. Since the discovery of the integer quantum Hall effect and the formalization of the Chern number, topological concepts in condensed matter physics have led researchers to identify a variety of non-trivial quantum materials, such as two-dimensional quantum spin Hall insulators \cite{kane2005quantum,bernevig2006quantum,roth2009nonlocal} with surface conduction channels locked to the spin of electric carriers, time-reversal symmetry-protected three-dimensional topological insulators \cite{fu2007topological} characterized by $\mathbb{Z}_2$ invariants that host surface Dirac cones mediating surface conductivity, and topological gapless Weyl semimetals \cite{hosur2013recent,soluyanov2015type,yan2017topological,armitage2018weyl}, which are intrinsically connected to the chiral anomaly and exhibit surface Fermi arcs that facilitate chiral transport.

While the Berry curvature is conceptually and functionally significant, it captures only part of the geometric properties of quantum states. The full geometric description of quantum states is given by the quantum geometric tensor (QGT) \cite{torma2023essay}, a complex quantity in which the imaginary part corresponds to the Berry curvature. Recent research, especially following the discovery of flat-band superconductivity in twisted bilayer graphene \cite{cao2018unconventional,cao2018correlated}, has highlighted the role of the real part of the QGT, known as the quantum metric (QM), in understanding superconductivity in flat bands \cite{torma2022superconductivity,huhtinen2022revisiting}. The QM is also connected to Hall viscosity \cite{avron1995viscosity,haldane2009hall,read2011hall,hoyos2012hall,holder2019unified}, Drude weight \cite{resta2011insulating,marrazzo2019local,rigol2008drude}, the stabilization of quantum states \cite{morales2023pressure,shi2024adiabatic}, kinetic coefficients in the time-dependent Ginzburg-Landau equation \cite{iskin2023extracting}, and quantum-induced transport phenomena \cite{kaplan2024unification,gao2023quantum,gao2019nonreciprocal}. Earlier studies of superfluidity in topological flat bands, such as those by Peotta et al. \cite{peotta2015superfluidity}, revealed that the QM constrains superfluid weight (SW), setting a lower bound defined by the Chern number of topological flat bands. This work established a foundational link between superfluidity and the geometric properties of topological quantum states. Furthermore, the QM defines a quantum volume and forms an inequality with the Chern number in Chern insulators, providing deeper insights into the interplay between quantum geometry and topology \cite{ozawa2021relations}.

For degenerate quantum states, the QGT becomes a non-Abelian, matrix-valued quantity, with its real and imaginary parts defining the non-Abelian QM and Berry curvature, respectively \cite{ma}. The non-Abelian Berry curvature generalizes the Abelian Berry curvature to degenerate subspaces \cite{wilczek1984appearance}, and its interplay with quantized Wilson loops in higher-order topological insulators can induce Bloch oscillations \cite{di2020non}. The Abelian QM of single-particle wave functions in topological flat bands determines the stability of fractional Chern insulators \cite{wu2024quantum}, induces nonlinear transport in topological antiferromagnets \cite{wang2023quantum}, and relates to the SW \cite{peotta2015superfluidity}. Employing Pl\"ucker embeddings to represent arbitrary classifying spaces, Bouhon et al. enable the quantification of geometric properties in multi-band systems \cite{bouhon2023quantum}

In this work, we demonstrate that near the superconducting transition temperature, the SW includes an additional contribution, beyond the normal contributions, which is proportional to the trace of the non-Abelian QM. Using a concrete model with time-reversal symmetry and degenerate bands in the normal state, we show that, upon entering the superconducting state, the QM contribution becomes significant and can even dominate the SW under certain parameters, despite the Chern number being zero due to time-reversal symmetry. Furthermore, through ab initio DFT calculations, we identify $\text{MoS}_2$ and $\text{TiSe}_2$ as exceptional material platforms to experimentally probe the interplay between non-Abelian QM and superconductivity, establishing a solid foundation for future exploration of geometric effects in quantum materials.
 
\textit{Non-Abelian QM and SW}\textbf{---} The non-Abelian QM plays a crucial role in understanding the SW in systems with degenerate bands. Here, we review its definition and establish its connection to the SW, with the detailed derivation provided in Appendix \ref{appa}. 

Consider a Hamiltonian $H(\mathbf{k})$ that contains $N$ degenerate bands, with a finite gap separating them from the remaining bands. We denote the degenerate eigenenergies as $\epsilon_\mathbf{k}$ and the corresponding eigenstates as ${\mid u_{1,\mathbf{k}}\rangle, \dots, \mid u_{N,\mathbf{k}}\rangle }$. For a general state $\mid \Psi_\mathbf{k}\rangle \equiv \sum_{j=1}^{N} c_{j}\left(\mathbf{k}\right) \mid u_{j,\mathbf{k}}\rangle$ in the degenerate subspace, the distance between $\mid \Psi_\mathbf{k}\rangle$ and $\mid \Psi_{\mathbf{k} + d\mathbf{k}}\rangle$ can be expressed as \cite{ma,ding2024non,ding2022extracting,bouhon2023quantum}:
\begin{equation}
dS^2=1-|\langle\Psi_\mathbf{k}\mid \Psi_{\mathbf{k}+d\mathbf{k}}\rangle|^2=\sum_{\mu\nu} C^{\dagger}Q_{\mu\nu}C dk^{\mu}dk^{\nu},
\label{eq:qdistance}
\end{equation}
where $k^\mu$ represents the $\mu$-th component of the Bloch momentum in the Brillouin zone, $C = [c_1(\mathbf{k}), \dots, c_N(\mathbf{k})]^T$ is a column vector of complex coefficients, and the $N \times N$ matrix $Q_{\mu\nu}$ is the non-Abelian QGT, which is given by:

\begin{equation}
Q^{ij}_{\mu\nu}\equiv \langle \partial_\mu u_{i,\mathbf{k}}\mid \left(1-P_\mathbf{k}\right)\mid \partial_\nu u_{j,\mathbf{k}}\rangle.
\label{eq:QGT}
\end{equation}
Here, the projection operator is defined as $P_\mathbf{k} \equiv \sum_{j=1}^{N} \mid u_{j,\mathbf{k}}\rangle\langle u_{j,\mathbf{k}}\mid$, where the indices $i$ and $j$ label the states within the degenerate subspace. The QGT can be written as $Q_{\mu\nu} = R_{\mu\nu} - i\Omega_{\mu\nu}/2$, where $\Omega_{\mu\nu}$ and $R_{\mu\nu}$, which are antisymmetric and symmetric in the spatial indices $\mu$ and $\nu$, denote the non-Abelian Berry curvature and the non-Abelian QM, respectively.

The complementary projection operator, $1-P_\mathbf{k}$, ensures that the non-Abelian QGT is $U(N)$ gauge invariant, where $U(N)$ denotes the unitary transformation acting on the $N$-fold degenerate subspace. Notably, the trace of the non-Abelian QGT is given by: 
\begin{align}
\text{Tr} Q_{\mu\nu} &= \sum_{i=1}^{i=N}\langle \partial_{\mu} u_{i,\mathbf{k}} | (1 - |u_{i,\mathbf{k}}\rangle \langle u_{i,\mathbf{k}}|)| \partial_{\nu} u_{i,\mathbf{k}}  \rangle  \notag \\
&\quad - \sum_{i=1}^{i=N} \sum_{j=1,j\neq i}^{j=N}\langle\partial_{\mu} u_{i,\mathbf{k}}|u_{j,\mathbf{k}}\rangle\langle u_{j,\mathbf{k}}| \partial_{\nu} u_{i,\mathbf{k}}\rangle,
\label{QGT2}
\end{align}
where the first term corresponds to the summation of the Abelian QGT for each band, while the second term introduces additional inter-band contributions. As a result, the non-Abelian QGT is not simply the sum of the Abelian QGTs for individual bands.

To investigate the effect of the non-Abelian QM on the SW, we consider a system described by the Hamiltonian $H(\mathbf{k})$ with time-reversal symmetry. As the temperature decreases, the system is assumed to transition into an $s$-wave superconducting state. Assuming the Cooper pairs with momentum $\mathbf{q}$ and working in the band basis \cite{peotta2015superfluidity}, the BdG equation is given by:
\begin{equation}
H_{BdG}\left(\mathbf{k,q}\right)=\left[\begin{array}{cc}
\zeta\left(\mathbf{k+q}\right) & \hat{\Delta}\left(\mathbf{k},\mathbf{q}\right)\\
\hat{\Delta}^{\dagger}\left(\mathbf{k},\mathbf{q}\right) & -\zeta(\mathbf{k-q})
\end{array}\right]
\label{eq:bdg}
\end{equation}
where $\zeta(\mathbf{k+q})$ is a diagonal matrix whose components are the eigen-energies of the Hamiltonian $H(\mathbf{k+q})$, and the pairing potential matrix is denoted by $\hat{\Delta}(\mathbf{k},\mathbf{q})$. The $ij$-th component of $\hat{\Delta}(\mathbf{k},\mathbf{q})$ is given by $\hat{\Delta}^{ij}(\mathbf{k},\mathbf{q}) = \Delta\langle u_{i,\mathbf{k+q}} \mid \Theta u_{j,\mathbf{-k+q}} \rangle$, where $\Delta$ is the s-wave pairing strength and $\Theta$ is the time-reversal operator, satisfying $\Theta^2 = \pm1$. Given the finite gap between the degenerate subspace of interest and the other bands, we will focus on the degenerate subspace and investigate the corresponding superconductivity.

The SW reflects a material's capacity to sustain a dissipationless current, becoming nonzero below the critical temperature of the superconducting transition, thus defining superconductivity. The SW can be calculated via the second-order derivative of the free energy $F(\mathbf{q})$ with respect to the momentum of the Cooper pairs, $\mathbf{q}$, i.e., $\mathscr{D}_{\mu\nu} = \lim_{{\mathbf{q} \rightarrow \mathbf{0}}} \frac{\partial^{2}F(\mathbf{q})}{\partial q_{\mu} \partial q_{\nu}}$. At low temperatures, the superconductivity of the system can be fully captured by the degenerate subspace due to the energy gap. 

Near the superconducting transition temperature, where the pairing strength $\Delta$ is small compared to the temperature $k_B T$, and condition $|\hbar \mathbf{q}| \ll |\Delta/T|$ is assumed (with the Boltzmann constant $k_B=1$ and Planck constant $\hbar=1$ set to zero throughout this work), we leverage the properties of the sewing matrix $B_{\mathbf{k}}^{ij} \equiv \langle u_{i,\mathbf{-k}} \mid \Theta u_{j,\mathbf{k}} \rangle$, which serves as a key tool for characterizing the $\mathbb{Z}_2$ topological invariant \cite{bernevig2013topological}. Here, $i$ and $j$ label the band indices in the degenerate subspace, and $\Theta$ denotes the time-reversal symmetry operator. 

The free energy near the transition temperature  can be approximated as follows (for details, refer to the Appendix \ref{appa}):

\begin{equation}
F\left(\mathbf{q}\right) \approx F^{0}+F^{N}+F^{QM}.
\label{eq:free}
\end{equation}
The term $F^{0}$ is independent on $\Delta$ and 
\begin{equation}
\label{eq:QMfree}
\begin{cases}
F^{N}  =\frac{\Delta^2}{g}+\Delta^{2}\frac{N}{2}\int\frac{d\mathbf{k}}{\left(2\pi\right)^{d}}\frac{n_{F}\left(\epsilon_{\mathbf{k+q}}\right)-n_{F}\left(-\epsilon_{-\mathbf{k+q}}\right)}{\epsilon_{\mathbf{k+q}}+\epsilon_{-\mathbf{k+q}}}\\
F^{QM}  =-2\Delta^{2}q_{\mu}q_{\nu}\int\frac{d\mathbf{k}}{\left(2\pi\right)^{d}}\frac{\sum_{n=1}^{n=N}R_{\mu\nu}^{nn}\left(\mathbf{k}\right)\left[n_{F}\left(\epsilon_{\mathbf{k}}\right)-n_{F}\left(-\epsilon_{-\mathbf{k}}\right)\right]}{\epsilon_{\mathbf{k}}+\epsilon_{-\mathbf{k}}},
\end{cases}
\end{equation}
where $n_{F}\left(x\right)$ denote the Fermi-Dirac distribution. As shown in the second equation of Eq. (\ref{eq:QMfree}), the QM $R_{\mu\nu}^{nn}$ contributes to the free energy, which in turn affects the SW. The full SW is expressed as::
\begin{equation}
\mathscr{D}_{\mu\nu}=\mathscr{D}^N_{\mu\nu}+\mathscr{D}^{QM}_{\mu\nu},
\label{QM}
\end{equation}
the term $\mathscr{D}^{N}_{\mu\nu}\equiv \lim_{{\mathbf{q} \rightarrow \mathbf{0}}} \frac{\partial^{2}F^N(\mathbf{q})}{\partial q_{\mu} \partial q_{\nu}}$ and the term

\begin{equation}
\mathscr{D}^{QM}_{\mu\nu}=\int\frac{d\mathbf{k}}{\left(2\pi\right)^{d}}\frac{-2\Delta^{2}Tr \left[R_{\mu\nu}\left(\mathbf{k}\right)\right] \left[n_{F}\left(\epsilon_{\mathbf{k}}\right)-n_{F}\left(-\epsilon_{-\mathbf{k}}\right)\right]}{\epsilon_{\mathbf{k}}+\epsilon_{-\mathbf{k}}},
\label{QM1}
\end{equation}
where the trace acts on the degenerate subspace. Equation (\ref{QM1}) shows that the SW is proportional to the trace of the non-Abelian QM and, therefore, does not equal the summation of the SW of each band, as it originates from the definition of the non-Abelian QGT.

\textit{Lattice model and results}\textbf{---} Next, we consider the contribution of the non-Abelian QM to the SW in a specific model with the Hamiltonian
\begin{equation}
H\left(\mathbf{k}\right)=d_{1}\left(\mathbf{k}\right)\Gamma_{zx}+d_{2}\left(\mathbf{k}\right)\Gamma_{xx}+d_{3}\left(\mathbf{k}\right)\Gamma_{0z}-\mu,
\label{eq:hami}
\end{equation}
where $\mathbf{k}=\left(k_x,k_y\right)$ denotes the Bloch momentum, $\Gamma_{ij} \equiv \sigma_i \otimes s_j$ with $i,j \in {0,x,y,z}$, and $\sigma_i$ and $s_j$ represent Pauli matrices acting on the orbital and the (pseudo)spin spaces, respectively. The coefficients are $d_{1} \left(\mathbf{k}\right)= \sin k_{x}$, $d_{2} \left(\mathbf{k}\right)= -\sin k_{y}$, and $d_{3}\left(\mathbf{k}\right) = -\left(M - \cos k_{x} - \cos k_{y}\right)$. The Hamiltonian (Eq.\ref{eq:hami}) satisfies time-reversal symmetry, $\Theta H(\mathbf{k}) \Theta^{-1} = H(-\mathbf{k})$, where $\Theta = \sigma_0 \otimes s_z \mathcal{K}$ with $\mathcal{K}$ denoting complex conjugation. The Hamiltonian also satisfies inversion symmetry, $\mathcal{P}H(\mathbf{k})\mathcal{P}^{-1} = H(-\mathbf{k})$, and $PT$ symmetry, $\mathcal{PT}H(\mathbf{k})\mathcal{PT}^{-1} = H(\mathbf{k})$. Here, the symmetry operators are $\mathcal{P} \equiv \sigma_0 \otimes s_z$ and $\mathcal{PT} = \mathcal{K}$. 

\begin{figure}[h]
\includegraphics[width=1\columnwidth,height=1\textheight,keepaspectratio]{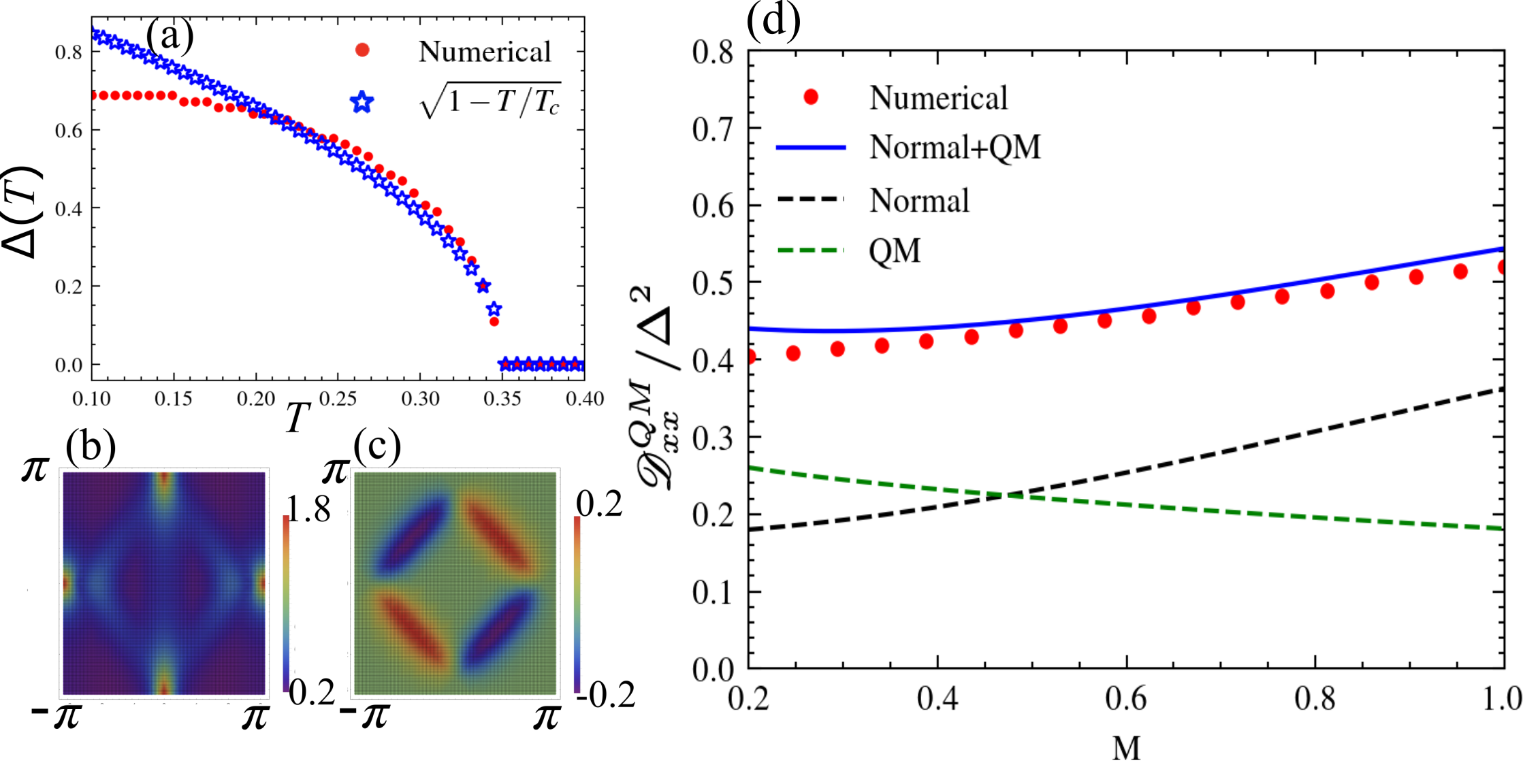}
\caption{Pairing potential  $\Delta(T)$ vs temperature $T$, QM and SW. (a) The red dots represent numerical results, while the blue stars show the temperature-dependent gap of a standard s-wave superconductor. The parameters are $g=1.57$, $M = 0.5$, $\mu = 2$, with the superconducting transition temperature $T_c \approx 0.35$. (b, c) The xx and xy components of the trace of the non-Abelian QM, where the trace is taken over the degenerate subspace of interest. (d) SW $\mathscr{D}^{QM}_{xx}$, in units of $\Delta^2$, as a function of $M$. The red dots are obtained via numerical calculation, while the normal (black dashed lines) and non-Abelian QM (green dashed lines) contributions to the SW are calculated using the terms $\mathscr{D}^N_{\mu\nu}$ and $\mathscr{D}^{QM}_{\mu\nu}$ in Eq. (\ref{QM}), respectively. The blue line represents the sum of the normal and non-Abelian QM contributions. }
\label{Fig.1}
\end{figure}

The system described by the Hamiltonian above exhibits a pair of twofold degenerate bands. Assuming $s$-wave pairing, the pairing potential is calculated self-consistently, revealing its temperature dependence, $T$, as illustrated in Fig. \ref{Fig.1}(a). This result indicates that the system undergoes a transition from the normal state to a superconducting state as the temperature decreases.

As shown in Eq. (\ref{QM}), the SW depends on the trace of the non-Abelian QM in the degenerate subspace. The trace of the non-Abelian QM obeys the following equation:
\begin{equation}
\mathrm{Tr}R_{\mu\nu}\left(\mathbf{k}\right)=\frac{1}{2}\partial_{\mu}\hat{\mathbf{d}}\cdot\partial_{\nu}\hat{\mathbf{d}},
\label{eq:degQM}
\end{equation} 
where the unit vector $\hat{\mathbf{d}} \equiv \frac{(d_1, d_2, d_3)}{|\mathbf{d}|}$. In a two-dimensional system, the trace of the non-Abelian QM $R_{xy}$ at any Bloch momentum $\mathbf{k}$ in the BZ can be canceled by $R_{xy}$ at another Bloch momentum $-\mathbf{k}$, as illustrated in Fig. \ref{Fig.1} (c). Therefore, based on Eq. (\ref{QM}) and the condition $\epsilon_{\mathbf{k}} = \epsilon_{-\mathbf{k}}$ for the time-reversal symmetric normal state, the SW $\mathscr{D}^{QM}_{xy} = \mathscr{D}^{QM}_{yx} = 0$. However, the non-Abelian QM $R_{xx}$ remains positive throughout the entire BZ, as shown in Fig. \ref{Fig.1} (b). Thus, the non-Abelian QM $R_{xx}$ can contribute a nonzero value to the SW $\mathscr{D}^{QM}_{xx}$.

To further investigate the importance of the non-Abelian QM in contributing to the SW, we obtain the SW numerically and compare it to the normal and non-Abelian contributions based on the analytical expression in Eq. (\ref{QM}) for different values of the parameter $M$. As shown in Fig.  \ref{Fig.1} (d), the contribution to the SW from non-Abelian QM is comparable to that of the normal term, highlighting the necessity of non-Abelian QM for accurately obtaining the correct SW. Notably, for $M < 0.55$, the contribution from non-Abelian QM exceeds that of the normal term.

\textit{Non-Abelian QM and SW in $\text{MoS}_2$ and TiSe$_2$}\textbf{---} Monolayer transition-metal dichalcogenides, such as $\text{MoS}_2$ and TiSe$_2$, have attracted considerable interest due to their unique electronic structures, which include isolated degenerate bands enabling non-Abelian quantum phenomena \cite{trolle2014theory}. Under gate-induced tuning, $\text{MoS}_2$ can exhibit superconducting states \cite{ye2012superconducting,costanzo2016gate}, while TiSe$_2$ hosts both charge density wave (CDW) and superconducting phases at low temperatures. The CDW phase, however, can be suppressed by intercalation with Cu or Li, or through the application of pressure \cite{liao2021coexistence,morosan2006superconductivity,PhysRevLett.103.236401}, positioning these materials as promising platforms for exploring the influence of the non-Abelian QM on superfluid weight.

\begin{figure}[h]
\includegraphics[width=1.\columnwidth,height=1.\textheight,keepaspectratio]{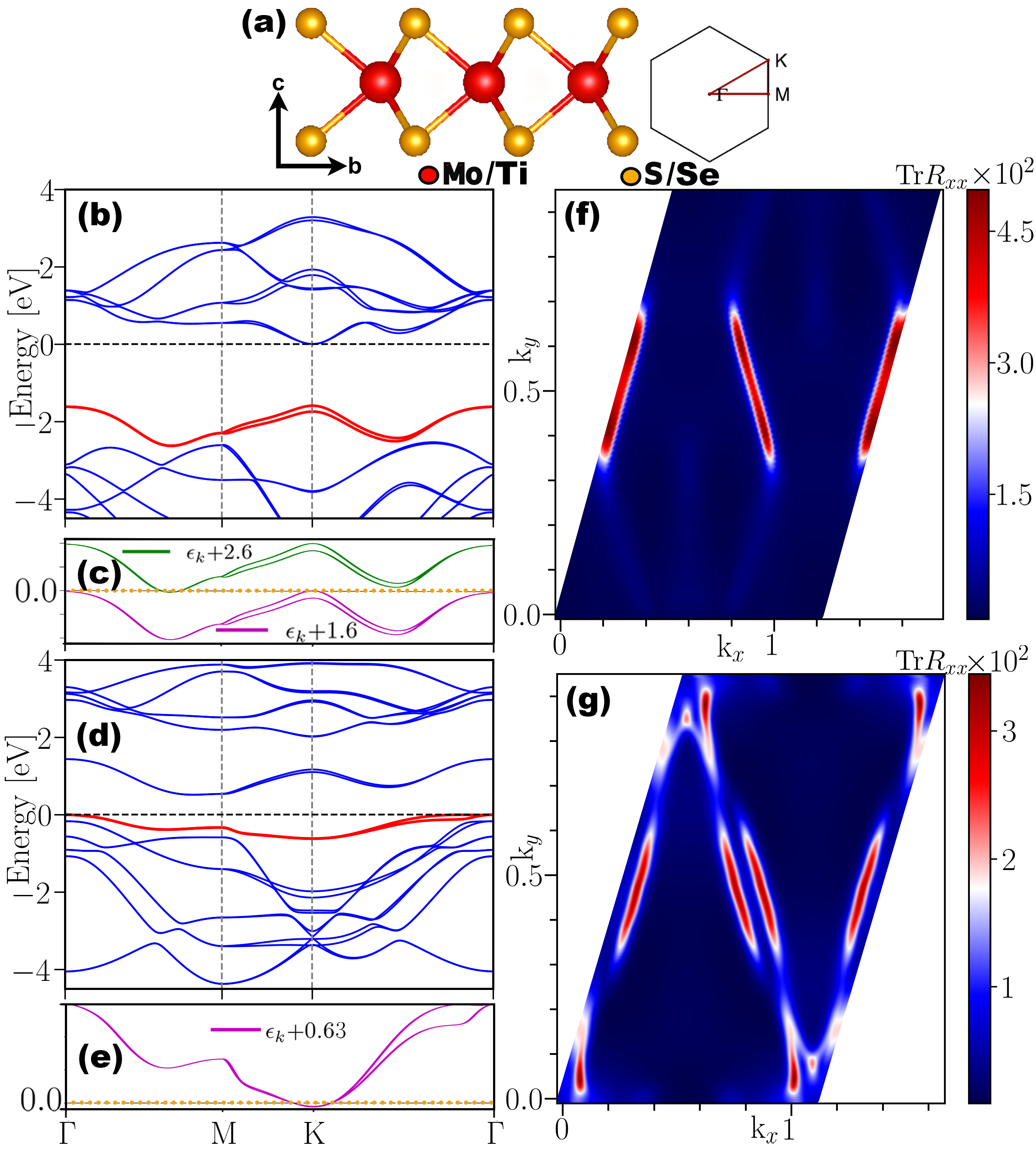}
\caption{DFT calculations of band structures and non-Abelian QM. (a) The left panel shows a single layer of MoS$_2$ or TiSe$_2$, and the right panel displays their BZ. (b, d) Band structures of MoS$_2$ and TiSe$_2$ derived from the Wannier Hamiltonian. (c, e) Shifts in the targeted bands due to hole doping. (f, g) Distribution of the trace of the $xx$ component of the non-Abelian QM in the BZ, contributed by the targeted bands for MoS$_2$ and TiSe$_2$, respectively.}
\label{fig4}
\end{figure}

To investigate the non-Abelian QM and SW in $\text{MoS}_2$ and TiSe$_2$, we constructed Wannier Hamiltonians for each material. For $\text{MoS}_2$, the Wannier basis includes Mo:4$d$ and S:3$p$ orbitals, while for $\text{TiSe}_2$, it includes Ti/Se:3$d$/4$p$ orbitals. These Hamiltonians demonstrate excellent agreement with DFT-calculated bands within the energy range of -8 to 3 eV. The two valence bands closest to the Fermi level are isolated and nearly degenerate, as highlighted in red in Fig. \ref{fig4}(b, d). This study focuses on the non-Abelian QM and the associated SW of these bands. Notably, the trace of the non-Abelian QM, $R_{xx}$, for these bands shows a nonzero distribution across the BZ, as depicted in Fig. \ref{fig4}(f, g). 

\begin{figure}[h]
\includegraphics[width=1.\columnwidth,height=1.\textheight,keepaspectratio]{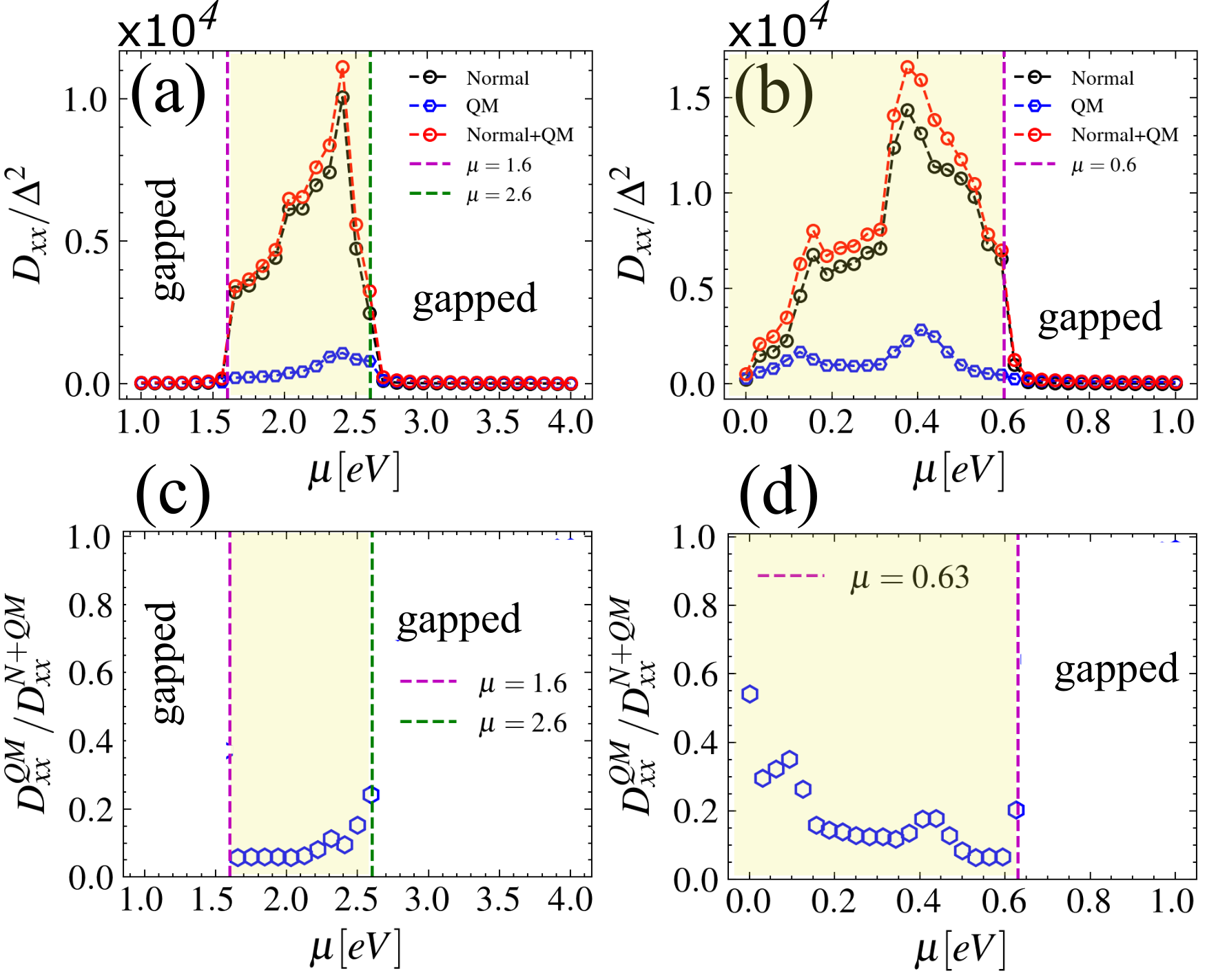}

\caption{ SWs based on DFT calculations. (a, b) Contributions of the normal and non-Abelian QM to the SWs $\mathscr{D}^{QM}_{xx}$, in units of $\Delta^2$, as a function of chemical potential $\mu$ for MoS$_2$ and TiSe$_2$, respectively. (c, d) Proportion of the SW contributed by the QM component relative to the total SW for MoS$_2$ and TiSe$_2$, respectively. The gapped regions corresponding to the normal phase are in insulating phases.}
\label{figr}
\end{figure}

Using the theoretical formula in Eq. (\ref{QM1}), which connects the non-Abelian QM and band properties to the SW, we compute the SW within the Wannier Hamiltonian framework, as shown in Fig. \ref{figr}. For $\text{MoS}_2$ (Fig. \ref{figr}(a)), the non-Abelian QM contribution to the SW becomes significant when $\text{MoS}_2$ transitions from its normal phase to a metallic state. This transition corresponds to the slab region, with boundaries marked by magenta and green dashed lines. Outside this region, the normal phase is gapped. The non-Abelian QM contribution to the total SW reaches approximately $20\%$.

Similar results are found in $\text{TiSe}_2$ (Figs. \ref{figr}(b, d)), where the non-Abelian QM contribution to the total SW is about $50\%$. These findings highlight that the non-Abelian QM effect on SW in $\text{MoS}_2$ and $\text{TiSe}_2$ is substantial and cannot be neglected.

DFT calculations were performed using the full-potential local orbital code, version 22.00-62 \cite{koepernik1999full}. The exchange-correlation potential was treated with the generalized gradient approximation formulated by Perdew, Burke, and Ernzerhof \cite{perdew1996generalized}. Reciprocal space integrations employed the linear tetrahedron method with Bl\"ochl corrections, using a $12 \times 12 \times 1$ k-point grid for both self-consistent calculations and band structure analysis within the BZ.

The localized Wannier basis was constructed using Mo/Ti $4d$/$3d$ and S/Se $3p$/$4p$ orbitals, utilizing the same $k$-mesh as the self-consistent calculations to build the Wannier model. This Wannier Hamiltonian was subsequently used to compute $\text{Tr } R_{xx}$ and the SWs. The lattice parameters were set to $a = b = 3.1428$ \AA{} for MoS$_2$ and $a = b = 3.4953$ \AA{} for TiSe$_2$. To ensure monolayer behavior and eliminate interlayer interactions, a vacuum layer of 20 \AA{} was added along the $z$ direction.

\textit{Conclusion}\textbf{---} In conclusion, we have elucidated the significant role of the non-Abelian QM in influencing the SW in time-reversal symmetric systems. Our results demonstrate that near the superconducting transition temperature, the SW comprises contributions from both normal and QM terms, with the non-Abelian QM proving particularly influential.

We first demonstrate that near the superconducting transition temperature, the SW is proportional to the non-Abelian QM in general systems with degenerate bands and time-reversal symmetry. By studying a lattice model with a time-reversal symmetric Hamiltonian containing a pair of degenerate bands, we find that the contribution of the non-Abelian QM to the SW can be comparable to or even exceed the conventional term, highlighting the significant role of non-Abelian geometric effects in superconductivity. This result is further corroborated by \textit{ab initio} DFT calculations, which show that the non-Abelian QM contributes up to 20\% of the SW in MoS$_2$ and as much as 50\% in TiSe$_2$.

These findings not only enhance our understanding of the interplay between quantum geometry and superconductivity but also open avenues for exploring novel quantum materials where non-Abelian properties may be exploited. This work underscores the importance of geometric features in quantum states, suggesting that further investigations into non-Abelian effects could yield rich insights into the collective behaviors of topological systems.

\textit{Acknowledgments}\textbf{---} K.C., B.K. and P.H. acknowledge support from National Science Foundation grant no. DMR 2047193. 

\noindent $^{\chi}$ kaiqcatchen@gmail.com
\noindent $^{\dagger}$phosur@central.uh.edu

 \bibliographystyle{apsrev4-2}
\bibliography{nonab}

 \onecolumngrid

\section*{Appendix for The Effect of the Non-Abelian Quantum Metric on Superfluidity}
In this supplementary material, we provide details on the derivation of the connection between superfluid weight and the non-Abelian QM.

\appendix 
\section{Building the connection between SW and non-Abelian QM }
Let's consider the time-reversal symmetric Hamiltonian $H\left(\mathbf{k}\right)$. The time-reversal symmetry relates the Bloch eigenstate $\mid u_{m,\mathbf{k}}\rangle$ and eigenenergy $\epsilon_{\mathbf{k}}$ at momentum $\mathbf{k}$ to the Bloch eigenstate $\Theta \mid u_{m,-\mathbf{k}}\rangle$ and eigenenergy $\epsilon_{-\mathbf{k}}$ at $\mathbf{-k}$, respectively. After writing the BdG Hamiltonian in the band basis, the pairing potential is related to the quantum geometric tensor. The off-diagonal components of the pairing potential in the band basis vanish, making the pairing potential block diagonal. Therefore, we can separately consider the contributions of each degenerate band to the SW. After setting the chemical potential close to the bands of interest and at lower temperatures (below the gap between the bands of interest and the remaining bands), physically relevant quantities, such as the SW, can be obtained by considering only the degenerate band with eigenenergy $\epsilon_{\mathbf{k}}$. The BdG Hamiltonian can be expressed as follows:
\begin{equation}
H_{BdG}\left(\mathbf{k,q}\right)=\left[\begin{array}{cc}
\zeta_\mathbf{k+q} & \hat{\Delta}\left(\mathbf{k},\mathbf{q}\right)\\
\hat{\Delta}^{\dagger}\left(\mathbf{k},\mathbf{q}\right) & -\zeta_\mathbf{k-q},
\end{array}\right]
\end{equation}
where $\zeta_{\mathbf{k+q}}$ is a diagonal matrix whose components are the eigenenergies of the Hamiltonian $H(\mathbf{k+q})$. As stated above, we only consider the degenerate subspace; hence, $\zeta_{\mathbf{k+q}} = \epsilon_{\mathbf{k+q}} I_{N \times N}$, and the pairing potential matrix is denoted by $\hat{\Delta}(\mathbf{k}, \mathbf{q})$. The $ij$-th component of $\hat{\Delta}(\mathbf{k},\mathbf{q})$ is given by $\hat{\Delta}^{ij}(\mathbf{k},\mathbf{q}) = \Delta\langle u_{i,\mathbf{k+q}} \mid \Theta u_{j,\mathbf{-k+q}} \rangle$, where $\Delta$ is the s-wave pairing strength and $\Theta$ is the time-reversal operator. 

To obtain the SW, let's first derive the free energy, which is given by:
\begin{equation}
F=\frac{\Delta^2}{g}-\frac{1}{2\beta}\sum_{n}\int\frac{d\mathbf{k}}{\left(2\pi\right)^{d}}\mathrm{Tr}\ln G^{-1}\left(i\omega_{n},\mathbf{k},\mathbf{q}\right),
\end{equation}
where the trace $\mathrm{Tr}$ acts on the particle-hole space and the degenerate band subspace. The parameter $g$ represents the electron-phonon interaction strength, $\beta = \frac{1}{k_B T}$ is the inverse temperature, with $k_B = 1$ in this work, and the Matsubara frequency is defined as $\omega_n \equiv \frac{\pi(2n+1)}{\beta}$. The Green's function $G\left(i\omega_{n},\mathbf{k},\mathbf{q}\right)$ takes the following form:

\begin{equation}
G\left(i\omega_{n},\mathbf{k},\mathbf{q}\right)=\left[\begin{array}{cc}
\left(i\omega_{n}-\epsilon_{\mathbf{k+q}}\right) & -\hat{\Delta}\left(\mathbf{k},\mathbf{q}\right)\\
-\hat{\Delta}^{\dagger}\left(\mathbf{k},\mathbf{q}\right) & \left(i\omega_{n}+\epsilon_{-\mathbf{k+q}}\right)
\end{array}\right]^{-1}\equiv \left[G_{0}\left(i\omega_{n},\mathbf{k},\mathbf{q}\right)^{-1}-\Gamma\left(\mathbf{k},\mathbf{q}\right)\right]^{-1},
\end{equation}
in which the Green's function $G_{0}\left(i\omega_{n},\mathbf{k},\mathbf{q}\right)$ in the normal state and the pairing matrix $\Gamma\left(\mathbf{k},\mathbf{q}\right)$ are defined as follows:
\begin{equation}
G_{0}\left(i\omega_{n},\mathbf{k},\mathbf{q}\right)=\left[\begin{array}{cc}
\left(i\omega_{n}-\epsilon_{\mathbf{k+q}}\right)^{-1} & 0\\
0 & \left(i\omega_{n}+\epsilon_{\mathbf{-k+q}}\right)^{-1}
\end{array}\right],
\end{equation}
and 
\begin{equation}
\Gamma\left(\mathbf{k},\mathbf{q}\right)=\left[\begin{array}{cc}
0 & \hat{\Delta}\left(\mathbf{k},\mathbf{q}\right)\\
\hat{\Delta}^{\dagger}\left(\mathbf{k},\mathbf{q}\right) & 0
\end{array}\right].
\end{equation}
Near the superconducting critical temperature $T_{c}$, where the pairing potential $\hat{\Delta} \ll 1$, one can expand the free energy to second order in $\hat{\Delta}$ to obtain $F \approx F^{(0)} + F^{(2)} + \mathcal{O}\left(\hat{\Delta}^{4}\right)$, where $F^{(0)}$ is independent of $\hat{\Delta}$. The approximate free energy can then be expressed as:

\begin{equation}
F^{\left(2\right)}=\frac{\Delta^2}{g}+\frac{1}{4\beta}\sum_{n}\int\frac{d\mathbf{k}}{\left(2\pi\right)^{d}}\mathrm{Tr}\left[G_{0}\Gamma G_{0}\Gamma\right]=\frac{\Delta^2}{g}+\frac{1}{2\beta}\sum_{n}\int\frac{d\mathbf{k}}{\left(2\pi\right)^{d}}\frac{Tr\hat{\Delta}^{\dagger}\left(\mathbf{k,q}\right)\hat{\Delta}\left(\mathbf{k,q}\right)}{\left(i\omega_{n}-\epsilon_{\mathbf{k+q}}\right)\left(i\omega_{n}+\epsilon_{-\mathbf{k+q}}\right)}.
\end{equation}
in which the trace acts on the degenerate subspace. Employing the standard equality $\frac{1}{\beta}\sum_{n}h\left(\omega_{n}\right)=\frac{1}{2\pi i}\varointctrclockwise dz h\left(-iz\right) n_{F}\left(z\right)$ with the Fermi-Dirac distribution $n_{F}\left(z\right)=\left[\exp\left(\beta z\right)+1\right]^{-1}$, the equation above can be written as:
\begin{equation}
F\approx \frac{\Delta^2}{g}+\frac{1}{2}\int\frac{d\mathbf{k}}{\left(2\pi\right)^{d}}Tr\hat{\Delta}^{\dagger}\left(\mathbf{k,q}\right)\hat{\Delta}\left(\mathbf{k,q}\right)\frac{n_{F}\left(\epsilon_{\mathbf{k+q}}\right)-n_{F}\left(-\epsilon_{-\mathbf{k+q}}\right)}{\epsilon_{\mathbf{k+q}}+\epsilon_{-\mathbf{k+q}}}.
\label{sfree}
\end{equation}

In the following, we will show that the term $Tr\hat{\Delta}^{\dagger}\left(\mathbf{k,q}\right)\hat{\Delta}\left(\mathbf{k,q}\right)$ in Eq. (\ref{sfree}) is connected with the non-Abelian QM, i.e.,
\begin{equation}
Tr\hat{\Delta}^{\dagger}\left(\mathbf{k,q}\right)\hat{\Delta}\left(\mathbf{k,q}\right)=\Delta^{2} N-4\Delta^{2}q^{\mu}q^{\nu} Tr R_{\mu\nu}\left(\mathbf{k}\right),
\label{tdelta}
\end{equation}
where the trace $\mathrm{Tr}$ acts on the degenerate subspace with dimension $N$, $R_{\mu\nu}$ is the non-Abelian QM, and repeated indices are assumed to be summed over. 

In terms of the free energy, the SW can be represented as:
\begin{align}
\mathscr{D}_{\mu\nu} &= \mathscr{D}_{\mu\nu}^{N} + \mathscr{D}_{\mu\nu}^{QM} \equiv \frac{\partial^{2} F^{N}}{\partial q_{\mu} \partial q_{\nu}} \bigg|_{\mathbf{q} \rightarrow \mathbf{0}} + \frac{\partial^{2} F^{QM}}{\partial q_{\mu} \partial q_{\nu}} \bigg|_{\mathbf{q} \rightarrow \mathbf{0}} \nonumber\\
&= \frac{\partial^{2} F^{N}}{\partial q_{\mu} \partial q_{\nu}} \bigg|_{\mathbf{q} \rightarrow \mathbf{0}} - \Delta^{2} \int \frac{d\mathbf{k}}{(2\pi)^{d}} \left[\frac{n_{F}\left(\epsilon_{\mathbf{k}}\right) - n_{F}\left(-\epsilon_{\mathbf{k}}\right)}{\epsilon_{\mathbf{k}}} \mathrm{Tr} R_{\mu\nu}\left(\mathbf{k}\right)\right]
\end{align}
where 

\begin{equation}
F^{N}=\frac{\Delta^2}{g}+\Delta^{2}\frac{N}{2}\int\frac{d\mathbf{k}}{\left(2\pi\right)^{d}}\frac{n_{F}\left(\epsilon_{\mathbf{k+q}}\right)-n_{F}\left(-\epsilon_{-\mathbf{k+q}}\right)}{\epsilon_{\mathbf{k+q}}+\epsilon_{-\mathbf{k+q}}}.
\end{equation}
Now, we provide details on the derivation of Eq. (\ref{tdelta}). Assuming a small vector $\mathbf{q}$, the components of the pairing potential satisfy:

\begin{align}
\hat{\Delta}\left(-\mathbf{k,q}\right)\approx B_{\mathbf{k}}-q^{\mu}\left[\partial_{\mu}B_{\mathbf{k}}+2i\mathcal{A}_{\mu}\left(-\mathbf{k}\right)B_{\mathbf{k}}\right]+q^{\mu}q^{\nu}\left[2i\mathcal{A}_{\mu}\left(-\mathbf{k}\right)\partial_{\nu}B_{\mathbf{k}}-2\langle\partial_{\mu}u_{-\mathbf{k}}\mid\partial_{\nu}u_{\mathbf{-k}}\rangle B_{\mathbf{k}}+\frac{1}{2}\partial_{\mu}\partial_{\nu}B_{\mathbf{k}}\right]
\end{align}
and 
\begin{align}
\hat{\Delta}\left(-\mathbf{k,q}\right)^{\dagger}\approx B_{\mathbf{k}}^{\dagger}-q^{\mu}\left[\partial_{\mu}B_{\mathbf{k}}^{\dagger}-2iB_{\mathbf{k}}^{\dagger}\mathcal{A^{\dagger}}_{\mu}\left(-\mathbf{k}\right)\right]+q^{\mu}q^{\nu}\left[-2i\partial_{\nu}B_{\mathbf{k}}^{\dagger}\mathcal{A^{\dagger}}_{\mu}\left(-\mathbf{k}\right)-2B_{\mathbf{k}}^{\dagger}\langle\partial_{\mu}u_{-\mathbf{k}}\mid\partial_{\nu}u_{\mathbf{-k}}\rangle+\frac{1}{2}\partial_{\mu}\partial_{\nu}B_{\mathbf{k}}^{\dagger}\right]
\end{align}
where $\mathcal{A}_{\mu}^{mn}\left(\mathbf{k}\right) \equiv i\langle u_{m,\mathbf{k}} \mid \partial_{\mu} u_{n,\mathbf{k}} \rangle$ is the $mn$-th component of the non-Abelian Berry connection, and $B_{\mathbf{k}}^{nm} \equiv \langle u_{n,\mathbf{-k}} \mid \Theta u_{m,\mathbf{k}} \rangle$ is the $nm$-th component of the sewing matrix. Expanding $\hat{\Delta}^{\dagger}\left(-\mathbf{k,q}\right)\hat{\Delta}\left(-\mathbf{k,q}\right)$ to second order in $\mathbf{q}$, we obtain:

\begin{align}
\hat{\Delta}^{\dagger}\left(-\mathbf{k,q}\right)\hat{\Delta}\left(-\mathbf{k,q}\right)&=I-q^{\mu}\left[B_{\mathbf{k}}^{\dagger}\partial_{\mu}B_{\mathbf{k}}+2iB_{\mathbf{k}}^{\dagger}\mathcal{A}_{\mu}\left(-\mathbf{k}\right)B_{\mathbf{k}}\right]-q^{\mu}\left[\partial_{\mu}B_{\mathbf{k}}^{\dagger}B_{\mathbf{k}}-2iB_{\mathbf{k}}^{\dagger}\mathcal{A^{\dagger}}_{\mu}\left(-\mathbf{k}\right)B_{\mathbf{k}}\right]\nonumber\\
&+q^{\mu}q^{\nu}\left[2iB_{\mathbf{k}}^{\dagger}\mathcal{A}_{\mu}\left(-\mathbf{k}\right)\partial_{\nu}B_{\mathbf{k}}-2B_{\mathbf{k}}^{\dagger}\langle\partial_{\mu}u_{-\mathbf{k}}\mid\partial_{\nu}u_{\mathbf{-k}}\rangle B_{\mathbf{k}}+\frac{1}{2}B_{\mathbf{k}}^{\dagger}\partial_{\mu}\partial_{\nu}B_{\mathbf{k}}\right]\nonumber\\
&+q^{\mu}q^{\nu}\left[\partial_{\mu}B_{\mathbf{k}}^{\dagger}-2iB_{\mathbf{k}}^{\dagger}\mathcal{A^{\dagger}}_{\mu}\left(-\mathbf{k}\right)\right]\left[\partial_{\nu}B_{\mathbf{k}}+2i\mathcal{A}_{\nu}\left(-\mathbf{k}\right)B_{\mathbf{k}}\right]\nonumber\\
&+q^{\mu}q^{\nu}\left[-2i\partial_{\nu}B_{\mathbf{k}}^{\dagger}\mathcal{A^{\dagger}}_{\mu}\left(-\mathbf{k}\right)B_{\mathbf{k}}-2B_{\mathbf{k}}^{\dagger}\langle\partial_{\mu}u_{-\mathbf{k}}\mid\partial_{\nu}u_{\mathbf{-k}}\rangle B_{\mathbf{k}}+\frac{1}{2}\left(\partial_{\mu}\partial_{\nu}B_{\mathbf{k}}^{\dagger}\right)B_{\mathbf{k}}\right]
\equiv Q_{0}+Q_{1}+Q_{2}
\end{align}

where $Q_i$ denotes the $i$th order in $\mathbf{q}$, with $Q_0 = I_{N\times N}$ and 

\begin{align}
Q_{1}&=-q^{\mu}\left[B_{\mathbf{k}}^{\dagger}\partial_{\mu}B_{\mathbf{k}}\cancel{+2iB_{\mathbf{k}}^{\dagger}\mathcal{A}_{\mu}\left(-\mathbf{k}\right)B_{\mathbf{k}}}+\partial_{\mu}B_{\mathbf{k}}^{\dagger}B_{\mathbf{k}}\cancel{-2iB_{\mathbf{k}}^{\dagger}\mathcal{A^{\dagger}}_{\mu}\left(-\mathbf{k}\right)B_{\mathbf{k}}}\right]=-q^{\mu}\left[B_{\mathbf{k}}^{\dagger}\partial_{\mu}B_{\mathbf{k}}+\partial_{\mu}B_{\mathbf{k}}^{\dagger}B_{\mathbf{k}}\right]=0
\end{align}

\begin{align}
Q_{2} &= q^{\mu} q^{\nu} \left\{ \cancel{2iB_{\mathbf{k}}^{\dagger} \mathcal{A}_{\mu}\left(-\mathbf{k}\right) \partial_{\nu} B_{\mathbf{k}}} - 2B_{\mathbf{k}}^{\dagger} \langle \partial_{\mu} u_{-\mathbf{k}} \mid \partial_{\nu} u_{\mathbf{-k}} \rangle B_{\mathbf{k}} + \frac{1}{2} B_{\mathbf{k}}^{\dagger} \partial_{\mu} \partial_{\nu} B_{\mathbf{k}} + \partial_{\mu} B_{\mathbf{k}}^{\dagger} \partial_{\nu} B_{\mathbf{k}} \right. \nonumber \\
& \quad \left. \cancel{-2iB_{\mathbf{k}}^{\dagger} \mathcal{A^{\dagger}}_{\mu}\left(-\mathbf{k}\right) \partial_{\nu} B_{\mathbf{k}}} + \cancel{\partial_{\mu} B_{\mathbf{k}}^{\dagger} 2i \mathcal{A}_{\nu}\left(-\mathbf{k}\right) B_{\mathbf{k}}} + 4B_{\mathbf{k}}^{\dagger} \mathcal{A^{\dagger}}_{\mu}\left(-\mathbf{k}\right) \mathcal{A}_{\nu}\left(-\mathbf{k}\right) B_{\mathbf{k}} \right. \nonumber \\
& \quad \left. \cancel{-2i \partial_{\nu} B_{\mathbf{k}}^{\dagger} \mathcal{A^{\dagger}}_{\mu}\left(-\mathbf{k}\right) B_{\mathbf{k}}} - 2B_{\mathbf{k}}^{\dagger} \langle \partial_{\mu} u_{-\mathbf{k}} \mid \partial_{\nu} u_{\mathbf{-k}} \rangle B_{\mathbf{k}} + \frac{1}{2} \partial_{\mu} \partial_{\nu} B_{\mathbf{k}}^{\dagger} B_{\mathbf{k}} \right\} \nonumber \\
&= q^{\mu} q^{\nu} \left\{ -4B_{\mathbf{k}}^{\dagger} \langle \partial_{\mu} u_{-\mathbf{k}} \mid \partial_{\nu} u_{\mathbf{-k}} \rangle B_{\mathbf{k}} + 4B_{\mathbf{k}}^{\dagger} \mathcal{A^{\dagger}}_{\mu}\left(-\mathbf{k}\right) \mathcal{A}_{\nu}\left(-\mathbf{k}\right) B_{\mathbf{k}} \right. \nonumber \\
& \quad \left. + \frac{1}{2} \left[ \partial_{\mu} \left(B_{\mathbf{k}}^{\dagger} \partial_{\nu} B_{\mathbf{k}}\right) - \cancel{\left(\partial_{\mu} B_{\mathbf{k}}^{\dagger}\right) \partial_{\nu} B_{\mathbf{k}}} \right] + \frac{1}{2} \left[ \partial_{\mu} \left( \left(\partial_{\nu} B_{\mathbf{k}}^{\dagger}\right) B_{\mathbf{k}} \right) - \cancel{\left(\partial_{\nu} B_{\mathbf{k}}^{\dagger}\right) \partial_{\mu} B_{\mathbf{k}}} \right] + \cancel{\partial_{\mu} B_{\mathbf{k}}^{\dagger} \partial_{\nu} B_{\mathbf{k}}} \right\} \nonumber \\
&= q^{\mu} q^{\nu} \left\{ -4B_{\mathbf{k}}^{\dagger} \langle \partial_{\mu} u_{-\mathbf{k}} \mid \partial_{\nu} u_{\mathbf{-k}} \rangle B_{\mathbf{k}} + 4B_{\mathbf{k}}^{\dagger} \mathcal{A^{\dagger}}_{\mu}\left(-\mathbf{k}\right) \mathcal{A}_{\nu}\left(-\mathbf{k}\right) B_{\mathbf{k}} + \frac{1}{2} \left[ \partial_{\mu} \left( \cancel{B_{\mathbf{k}}^{\dagger} \partial_{\nu} B_{\mathbf{k}} + \left(\partial_{\nu} B_{\mathbf{k}}^{\dagger}\right) B_{\mathbf{k}}} \right) \right] \right\},
\end{align}

where the sewing matrix $B_{\mathbf{k}}$ satisfies $B^\dagger_{\mathbf{k}} B_{\mathbf{k}} = 1$, and $\langle\partial_{\mu}u_{-\mathbf{k}}\mid\partial_{\nu}u_{\mathbf{-k}}\rangle$ is understood as a matrix with indices corresponding to the band indices in the degenerate subspace.

Finally, we obtain
\begin{align}
Tr\hat{\Delta}^{\dagger}\left(-\mathbf{k,q}\right)\hat{\Delta}\left(-\mathbf{k,q}\right)&=N-4q^{\mu}q^{\nu}\sum_{n\in deg}\langle\partial_{\mu}u_{n,-\mathbf{k}}\mid\left(I-\hat{P}_{-\mathbf{k}}\right)\mid\partial_{\nu}u_{n,\mathbf{-k}}\rangle\nonumber\\ 
&=N-4q^{\mu}q^{\nu} Tr Q_{\mu\nu}=N-4q^{\mu}q^{\nu} Tr R_{\mu\nu},
\end{align}
where $deg$ denotes the set of band indices in the degenerate subspace. In the last equality, we use the fact that the imaginary part is antisymmetric under the exchange of indices $\mu$ and $\nu$.
\label{appa}
\end{document}